\documentclass[showkeys,twocolumn]{revtex4-1}
\usepackage{url}
\usepackage[T1]{fontenc}
\usepackage{graphicx}
\usepackage{amsfonts}
\usepackage{amsmath}
\usepackage{amssymb}
\usepackage{subfigure}

\newcommand{\C}{\mathbb{C}}
\newcommand{\bra}[1]{\langle{#1}|}
\newcommand{\ket}[1]{|{#1}\rangle} 
\newcommand{\1}{\mbox{$\mathbb{I}$}}

\newcommand{\ketbra}[2]{\ket{#1}\bra{#2}}
\newcommand{\braket}[2]{\bra{#2}{#1}\rangle}
\newcommand{\tr}{\mathrm{tr}}
\newcommand{\pmin}{p_\mathrm{min}}

\newcommand{\stackidx}[4]{
  \substack{
  #1  #2 \\
  #3  #4}
}

\begin{document}
\title{Noise effects in the quantum search algorithm from the computational complexity
point of view}
\author{Piotr Gawron}
\email{gawron@iitis.pl}
\author{Jerzy Klamka}
\author{Ryszard Winiarczyk}
\affiliation{Institute of Theoretical and Applied Informatics,\\
Polish Academy of Sciences,\\
ul.~Ba\l{}tycka 5, 44-100 Gliwice, Poland}

\begin{abstract}
We analyse the resilience of the quantum search algorithm in the presence of
quantum noise modelled as trace preserving completely positive maps. We study
the influence of noise on computational complexity of the quantum search
algorithm. We show that only for small amounts of noise the quantum search
algorithm is still more efficient than any classical algorithm.
\end{abstract}


\maketitle

\section{Introduction}
It is often said that the strength of quantum computation lies in the
phenomena of quantum superposition and quantum entanglement. These features of
quantum computation allow to perform the computation on all possible inputs that
fit to the quantum register. One of the greatest achievements in the theory of
quantum algorithms is quantum search algorithm introduced by Grover.
The detailed description of this algorithm can be found in \cite{Grover1996},
\cite{Grover1997}, \cite{Grover1998}, \cite{Bugajski2001}.

Any physical implementation of a~quantum computer will be error-prone  because
of the interaction of the computing device with the environment. In this paper
we investigate the resilience of Grover's algorithm in the presence of quantum
noise. We use the language of density matrices and quantum channels. Our goal is
to find the maximal amount of noise, for which the quantum algorithm is better,
in terms of mean number of operations, than classical algorithm. We aim to
achieve this objective by considering some classes of quantum channels modelling
environmentally induced noise.

The paper is organised as follows: in Section~\ref{sec:review} we make a short
review of the subject. In Section~\ref{sec:formalism} we describe the formalism
of quantum information theory. In Section~\ref{sec:overview} we present the
quantum search algorithm. In Section~\ref{sec:noisemodel} we introduce the noise
model we have applied to the system. In Section~\ref{sec:influence} we analyse the
results and finally in Section~\ref{sec:summary} we present some conclusions.

\section{Review of existing work}\label{sec:review}
The problem of influence of noise on the quantum search algorithm has been extensively
studied by various researchers. In \cite{Barnes1999} the authors discuss
the influence of classical field upon a~quantum system implementing Grover's
algorithm. The authors of the paper \cite{Pablo-Norman1999} ask similar question to
the one asked in this work, but use Gaussian noise model, which in their case is not
described in the language of quantum channels. In \cite{Long2000}
the authors analyse how imperfections in realizations of quantum gates influence the
probability of success of the quantum search algorithm. 
In \cite{Ellinas2001} the authors analyse the behaviour of the quantum 
search algorithm realized with the use of noisy $\pi/4$ rotation gates.

The effect of unitary noise
on the quantum search algorithm is studied in \cite{Shapira2003}. In
\cite{Shenvi2003} the authors examine the robustness of Grover's search algorithm to
a random phase error in the oracle and analyse the complexity of the search
process. In \cite{Azuma2005} the author studies decoherence in Grover's
quantum search algorithm using a~perturbative method. The authors of
\cite{Zhirov2005} use the methods of quantum trajectories to study the effects of
dissipative decoherence on the accuracy of the Grover's quantum search algorithm.
In \cite{Salas2007} the author numerically simulates Grover's algorithm
introducing random errors of two types: one- and two-qubit gate errors and
memory errors.

\section{Formalism of quantum information}
\label{sec:formalism}
\subsection{Dirac notation}
Throughout this paper we use Dirac notation. Symbol $\ket{\psi}$
denotes a~complex column vector, $\bra{\psi}$ denotes the row vector dual
to $\ket{\psi}$. The scalar product of vectors $\ket{\psi}$, $\ket{\phi}$ is denoted
by $\braket{\phi}{\psi}$. The outer product 
of these vectors is denoted as $\ketbra{\phi}{\psi}.$ Vectors are labelled in the
natural way: $\ket{0}:=\left(\begin{smallmatrix}1\\0\end{smallmatrix}\right)$, 
$\ket{1}:=\left(\begin{smallmatrix}0\\1\end{smallmatrix}\right)$. Notation like
$\ket{\phi\psi}$ denotes the tensor product of vectors and is equivalent to
$\ket{\phi}\otimes\ket{\psi}$.

\subsection{Density operators}
The most general state of a~quantum system is described by a~density operator.
In quantum mechanics a~density operator $\rho$ is defined as hermitian
($\rho=\rho^\dagger$) positive semi-definite ($\rho\geq 0$) trace one
($\tr{(\rho)}=1$) operator. When a basis is fixed the density operator can be
written in the form of a~matrix. Diagonal density matrices can be identified with
probability distributions, therefore this formalism is a~natural extension of
probability theory. 

Density operators are usually called quantum states. The set of quantum states
is convex \cite{Bengtsson2006} and its boundary consists of pure states
which in matrix terms are rank one projectors. Convex combinations of pure
states lie inside the set and are called mixed states.

\subsubsection{Entanglement}
Entanglement is one of the most important phenomena in quantum information theory. 
We say that state $\rho$ is separable iff it can be written in the
following form
\begin{equation}
\rho=\sum_{i=1}^{M} q_i\, \rho_i^A \otimes \rho_i^B,
\end{equation}
where $q_i>0$ and $\sum_{i=1}^{M} q_i=1$.
A state that is not separable is called entangled. It is an open problem of
great importance and under investigation, to decide if a~given quantum state is
entangled or not.

\subsubsection{Subsystems}
Given two states $\rho^A$, $\rho^B$ of two systems $A$ and $B$, the product state
$\rho^{AB}$ of the composed system is obtained by taking the Kronecker product
of the states i.e. $\rho^{AB}=\rho^A\otimes\rho^B.$

Let $[\rho^{AB}]_{kl}$ be a matrix representing a~quantum system composed of two
subsystems of dimensions $M$ and $N$. We want to index the matrix elements of
$\rho$ using two double indices $[\rho^{AB}]_{\stackidx{m}{\mu}{n}{\nu}},$ so that
Latin indices correspond to the system $A$ and Greek indices correspond to the
system $B$. The relation between indices is as follows $k=(m-1) N + \mu$, $l=(n-1) N
+ \nu$. 
The partial trace with respect to system $B$ reads
$\tr_B(\rho^{AB})=\sum_\mu \rho_{\stackidx{m}{\mu}{n}{\mu}}=\rho^A$, 
and the partial trace with respect to system $A$ reads
$\tr_A(\rho^{AB})=\sum_m \rho_{\stackidx{m}{\mu}{m}{\nu}}=\rho^B$.

Given the state of the composed system $\rho^{AB}$ the state of subsystems can
by found by the means of taking partial trace of $\rho^{AB}$ with respect to one
of the subsystems. It should noted that tracing-out is not a reversible
operation, so in a~general case
\begin{equation}
    \rho^{AB}\neq \tr_A(\rho^{AB})\otimes\tr_B(\rho^{AB}).
\end{equation}

\subsection{Completely positive trace-preserving maps (CPTP)}
We~say that an operation is physical if it transforms density operators into
density operators. Additionally we assume that physical operations are linear.
Therefore an operation $\Phi(\cdot)$ to be physical has to fulfil the following
set of conditions:
\begin{enumerate}
	\item For any operator $\rho$ its image under operation $\Phi$ has to have its
	trace and positivity preserved i.e. if 
	$\tr{(\rho)}=1, \rho\geq0, \rho=\rho^\dagger$ then
	$\tr{(\Phi(\rho))}=1, \Phi(\rho)\geq0, \Phi(\rho)=\Phi(\rho)^\dagger.$
	\item Operator $\Phi$ has to be linear:
	\begin{equation}
		\Phi\left(\sum_i p_i\rho_i \right)=\sum_i p_i \Phi\left(\rho_i \right).
	\end{equation}
	\item The extension of the operator $\Phi$ to any larger dimension that acts
	trivially on the extended system has to preserve positivity. 
	This feature is called complete positivity. It means that for all positive semi-definite
	$\rho,\xi\geq 0$ the following holds
	\begin{equation}
		(\Phi\otimes\1_{\dim{(\xi)}})\left(\rho\otimes\xi\right)
		=\Phi\left(\rho\right)\otimes\xi \geq 0.
	\end{equation}
\end{enumerate}
CPTP maps are often called quantum channels.

\subsubsection{Kraus form}
Any operator $\Phi$ that is completely positive and trace preserving can be
expressed in so called Kraus form \cite{Bengtsson2006}, which consists
of the finite set $\{E_k\}$ of Kraus operators -- matrices that fulfil the
completeness relation: $\sum_k {E_k}^\dagger E_k=\1$.
The image of state $\rho$ under 
the map $\Phi$ is given by 
\begin{equation}
    \Phi(\rho)=\sum_k E_k \rho {E_k}^\dagger.
\end{equation}

\subsection{Measurement}
Quantum states cannot be observed directly. In the literature one considers two
main types of measurements: Von Neumann measurement and POVM (Positive Operator
Valued Measure) measurement. In this paper we use only Von Neumann
measurement but for the sake of completeness we also define POVM measurement.

The mathematical formulation of Von Neumann measurement is given by a map from a
set of projection operators to real numbers. 

Let us consider an orthogonal complete set of projection operators $P=\{P_i\}_{i=1}^N$ and
the set of real measurement outcomes $O=\{o_i\}_{i=1}^N$. Mapping $P\rightarrow
O$ is called Von Neumann measurement. Assuming the system is in the state $\rho$,
the probability $p_i$ of measuring outcome $o_i$ is given by the relation $p_i=\tr(P_i
\rho)$.

POVM measurement can be considered as a generalisation of Von Neumann
measurement. Let us take a set of positive operators $F=\{F_i\}_{i=1}^N$ such
that $\sum_{i=1}^N F_i=\1$ and the set of real measurement outcomes
$O=\{o_i\}_{i=1}^N$. Mapping $F\rightarrow O$ is called POVM measurement. Given
the system is in the state $\rho$, the probability $p_i$ of measuring outcome $o_i$
is given by the relation $p_i=\tr(F_i \rho)$.

\section{Overview of the Grover's algorithm}
\label{sec:overview} 
Grover's unordered database search algorithm is one of the most important
quantum algorithms. This is due to the fact that many algorithmic problems can
be reduced to exhaustive search. 

The main idea of the algorithm is to amplify the probability of the state which
represents the sought element. The algorithm is probabilistic and may fail to
return the proper result. Fortunately the probability of success is reasonably
high.

\subsection{The problem}
Let $X$ be a~set and let $f: X\rightarrow\{0,1\}$, such that
\begin{equation}
    f(x)=
    \left\{
    \begin{array}{l}
        1 \Leftrightarrow x= x_0\\
        0 \Leftrightarrow x\neq x_0
    \end{array}
    \right., x\in X,
\end{equation}
for some marked $x_0\in X$.

For the sake of simplicity we assume that $X$ is a set of binary strings of length $n$.
Therefore $|X|=2^n$ and $f:\{0,1\}^n \rightarrow \{0,1\}$. We can map the set
$X$ to a~set of states over $\C^{\otimes 2^n}$ in the natural way: $x
\leftrightarrow \ket{x}$, forming orthogonal, complete set of vectors. The goal of
the algorithm is to find the marked element.

\subsection{The algorithm}
The Grover's algorithm is composed of two main procedures: the oracle and diffusion.

\subsubsection{Oracle}
By an oracle we call a function that marks one defined element. 
In the case of Grover's algorithm, the marking of the element is done by the negation 
of the amplitude of the sought state.

With the use of elementary quantum gates the oracle can be constructed using ancilla $\ket{q}$
in the following way:
\begin{equation}
\label{equ:oracle1}
   O\ket{x}\ket{q}=\ket{x}\ket{q\oplus f(x)},
\end{equation}
where $\oplus$ denotes addition modulo 2.
If the register $\ket{q}$ is prepared in the state
\begin{equation}
    \ket{q}=H\ket{1}=\frac{\ket{0}-\ket{1}}{\sqrt{2}},
\end{equation}
where $H$ denotes the Hadamard gate,
then, by substitution, Eq.~(\ref{equ:oracle1}) can be written as
\begin{equation}
    O\ket{x}\frac{\ket{0}-\ket{1}}{\sqrt{2}}=(-1)^{f(x)}\ket{x}\frac{\ket{0}-\ket{1}}{\sqrt{2}}.
\end{equation}
By tracing out the ancilla we get
\begin{equation}
    O\ket{x}=-(-1)^{f(x)}\ket{x}.
\end{equation}

\subsubsection{Diffusion}
The operator $D$ rotates any state around the state
\begin{equation}
\ket{\psi}= \frac{1}{\sqrt {2^n}}\sum\limits_{x = 0}^{2^n-1}\ket{x},
\end{equation}
where $D$ can be written as
\begin{equation}
D=-H^{\otimes n}(2\ketbra{0}{0}-\1)H^{\otimes
n}=2\ketbra{\psi}{\psi}-\1.
\end{equation}

\subsubsection{Initialisation}
We begin in the ground state $\ket{0\ldots 00}$. In the first step of the
algorithm we apply the Hadamard gate $H^{\otimes n}$ on the entire register.
This transforms the initial state into flat superposition of
computational base states:
\begin{equation}
H^{\otimes n}\ket{0\ldots 0}=
\frac{1}{\sqrt{n}}\left(\ket{0\ldots 00}+\ldots+\ket{1\ldots 11}\right).
\end{equation}

\subsubsection{Grover iteration}
The core of the algorithm consists of the applications of so called Grover
iteration gate $G=D\cdot O$. This procedure causes the sought 
state to be amplified and others states to be attenuated.

\subsubsection{Number of iterations}
The application of the diffusion operator on the base state $\ket{x}$ gives
\begin{equation}
D\ket{x}=-\ket{x_0}+\frac{2}{N}\sum_y\ket{y}.
\end{equation}
The application of this operator on any state gives
\begin{eqnarray*}
D\ket{x}&=&\sum_{i}\alpha_{i}( -\ket{x} +\frac{2}{N}y\sum_y\ket{y}) \\
&=&\sum_{i}( -\alpha_{i}+2s) \ket{x},
\end{eqnarray*}
where
\begin{equation}
s=\frac{1}{N}\sum_{i}\alpha _{i}.
\end{equation}

$k$-fold application of Grover's iteration $G$ on initial state $\ket{s}$ leads
to \cite{bouwmeester2000physics,Bugajski2001}
\begin{equation}\label{equ:grover1}
    G^k\ket{s} =\alpha_k\sum_{x\neq x_0}\ket{x}+\beta_k\ket{x_0},
\end{equation}
with real coefficients:
\begin{equation}\label{equ:grover2}
    \alpha _k=
    \frac{1}{\sqrt{N-1}} \cos\left(2k+1\right)
    \theta ,\quad \beta _k=\sin\left(2k+1\right) \theta,
\end{equation}
where $\theta$ is an angle that fulfils the relation
\begin{equation}
    \sin(\theta) =\frac{1}{\sqrt{N}}.
\end{equation}
Therefore the coefficients $\alpha_k,\beta_k$ are periodic functions of  $k$.
After the series of iterations $\beta_k$ rises. The influence of the marked
state $\ket{x_0}$ on the state of the register results in the evolution of the
initial state $\ket{s}$ towards the marked state.

The $\beta_k$ attains its maximum after approximately $\frac{\pi}{4}\sqrt{N}$ steps.
The number of steps needed to transfer the initial state towards the marked
state is of order $O(\sqrt{N})$. In the classical case the number of steps is of
order $O(N)$.

\subsubsection{Measurement} 
The last step of the Grover's algorithm is Von Neumann measurement.
The probability of obtaining the proper result is $|\beta_k|^2$.

\begin{figure*}[ht]
    \begin{center}
        \includegraphics[width=0.8\textwidth]{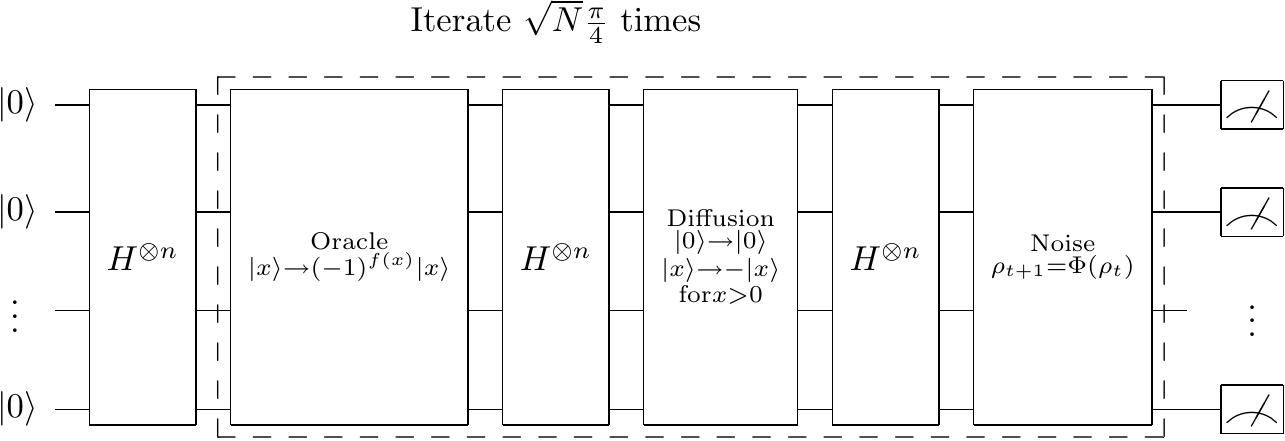}
    \end{center}
    \caption{The circuit for Grover's algorithm extended with a non-unitary noisy 
    channel.}
    \label{fig:grover}
\end{figure*}

\section{Noise model}
\label{sec:noisemodel}
The above discussion of quantum search algorithm has been conducted using state
vector formalism. In order to incorporate the noise into the quantum computation
model we need to make use of density operators which define the quantum state
in the most general way. 

\subsection{Quantum noise}
Microscopic systems that are governed by the laws of quantum mechanics are hard
to control and at the same time, to separate from the environment. The interaction
with the environment introduces noise into the quantum system. Therefore any 
future quantum computer will also be prone to noise.

\paragraph{One-qubit noise}\label{par:one-qubit-noise}
There are several one-parameter families of one-qubit noisy channels that are
typically discussed in the literature \cite{NC1999}. We present them
briefly below.

\subparagraph{Depolarising channel} 
This is a bi-stochastic channel that transforms any state into maximally mixed
state with a~given probability $\alpha$. The family of channels can be defined
using a~four-element set of Kraus operators
$$\left\{
\sqrt{1-\alpha}\1
,
\sqrt{\frac{\alpha}{3}}\sigma_x
,
\sqrt{\frac{\alpha}{3}}\sigma_y,
\sqrt{\frac{\alpha}{3}}\sigma_z
\right\},$$
where 
$$\1=
\left[
\begin{array}{ll}
1 & 0 \\ 
0 & 1
\end{array}
\right],
\sigma_x=
\left[
\begin{array}{ll}
0 & 1 \\ 
1 & 0
\end{array}
\right],
$$
$$
\sigma_y=
\left[
\begin{array}{ll}
0 & -i \\ 
i & 0
\end{array}
\right],
\sigma_z=
\left[
\begin{array}{ll}
1 & 0 \\ 
0 & -1
\end{array}
\right]
$$
are Pauli matrices.

\subparagraph{Amplitude damping} The amplitude damping channel transforms $\ket{1}$
into $\ket{0}$ with a~given probability $\alpha$. State $\ket{0}$ remains unchanged. 
The set of Kraus operators is following
$$\left\{
\left[
\begin{array}{cc}
1 & 0 \\
0 & \sqrt{1-\alpha}
\end{array}
\right]
,
\left[
\begin{array}{cc}
0 & \sqrt{\alpha} \\
0 & 0
\end{array}
\right]
\right\}.$$
\subparagraph{Phase damping}  Phase damping is purely quantum phenomenon which
describes the loss of quantum information without the loss of energy. It is
described by the following set of Kraus  operators
$$\left\{
\left[
\begin{array}{cc}
1 & 0 \\
0 & \sqrt{1-\alpha}
\end{array}
\right]
,
\left[
\begin{array}{cc}
0 & 0 \\
0 & \sqrt{\alpha}
\end{array}
\right]
\right\}.$$
\subparagraph{Bit flip} 
The bit flip family of channels is the quantum version of classical Binary
Symmetric Channel. The action of the channel might be interpreted in the
following way: it flips the state of a qubit from $\ket{0}$ to $\ket{1}$ and from
$\ket{1}$ to $\ket{0}$ with probability $\alpha$. Kraus operators for this
family of channels consist of a matrix proportional to the identity and 
a~matrix proportional to the negation gate
$$
\left\{
\sqrt{1-\alpha}\1,
\sqrt{\alpha}\sigma_x
\right\}.
$$
\subparagraph{Phase flip} The phase flip channel acts similarly to bit flip channel
with the distinction that $\sigma_z$ gate is applied randomly to the qubit
$$\left\{
\sqrt{1-\alpha}\1,
\sqrt{\alpha}\sigma_z
\right\}.
$$
\subparagraph{Bit-phase flip} The bit-phase flip channel may be considered as joint
application of bit and phase flip gates on a qubit. Its Kraus operators form
is as follows
$$\left\{
\sqrt{1-\alpha}\1
,
\sqrt{\alpha}\sigma_y
\right\}.
$$

In all the above families of channels the real parameter $\alpha\in [0,1]$ can be
interpreted as the amount of noise introduced by the channel. 

\paragraph{Multiqubit local channels}\label{par:multi-qubit-noise}
Our goal is to extend the noise acting on distinct qubits to the entire 
registers. We assume that the appearance of an error on a given qubit is independent
from an error appearing on any other qubits.

In order to apply noise operators to multiple qubits we form a new set of Kraus
operators acting on a larger Hilbert space.

We assume that we have the set of $n$ one-qubit Kraus operators $\{e_k\}_{k=1}^n$. We
construct the new set of $n^N$ operators $\{E_k\}_{k=1}^{n^N}$ that act on Hilbert space of
dimension $2^N$ by applying the following formula
\begin{equation}\label{equ:localchannel}
\{E_k\}=\bigcup_I \{e_{i_1}\otimes e_{i_2}\otimes\ldots \otimes e_{i_N}\},
\end{equation}
where $I=\{i_1\}_{i_1=1}^n\times\{i_2\}_{i_2=1}^n\times\ldots\times\{i_N\}_{i_N=1}^n.$

One should note that the extended channel $\Phi(\rho)=\sum_k E_k\rho
E_k^\dagger$ is by the definition local \cite{Bengtsson2006}.

By applying Eq.~(\ref{equ:localchannel}) to the sets of operators listed above
we obtain one-parameter families of local noisy channels, which we use in
further investigations.

\subsection{Application of noise to the algorithm}
In order to simulate noisy behaviour of the system implementing the algorithm we
apply a noisy channel after every Grover iteration. The evolution of the system is 
described by the following procedure, which is graphically depicted in Fig.~\ref{fig:grover}
\begin{enumerate}
	\item Prepare system in state $\rho_0:=\ketbra{0^{\otimes n}}{0^{\otimes n}}$.
	\item $\rho:=H^{\otimes n}\rho_0 H^{\otimes n \dagger}$
	\item $\lfloor\frac{\pi}{4}\sqrt{N}\rfloor$ times do:
	\begin{enumerate}
		\item apply Grover iteration $\rho:=G\rho G^\dagger$,
		\item apply noise $\rho:=\Phi(\rho)$.
	\end{enumerate}
	\item Perform orthogonal measurement in computational basis.
	The probability of finding the sought element $\xi$ is $p=\bra{\xi}\rho\ket{\xi}$.
\end{enumerate}

This approach simplifies the physical reality  but it is sufficient to study the
robustness of the algorithm in the presence of noise. In order to study the
discussed problem we make use of the numerical simulation. Therefore some
simplification is necessary as the size of the problem grows exponentially fast
with the number of qubits.

The tool we use is \texttt{quantum-octave} \cite{GawronEtAll2010}, a~library that 
contains functions for simulation and analysis of quantum processes.

In our model we assume that it is easy to verify if  the correctness of the
quantum search algorithm the result of quantum search results. It is an
assumption usually made in  the complexity analysis of search algorithms. 

\section{Analysis of the influence of noise on the efficiency of the algorithm}
\label{sec:influence}
An interesting question arises: ``What is the maximal amount of noise for
which Grover's algorithm is more efficient than any classical search algorithm?''

Grover's algorithm is probabilistic, therefore we cannot expect to obtain
a~valid outcome with certainty. We assume that if algorithm fails in a~given run
we will rerun it. There is a~certain number of~reruns for which quantum
algorithm is worse than classical. We are interested only in the statistical 
behaviour of algorithm and calculate the mean value of repetitions. 

Let $k=\lfloor\frac{N}{2}/\frac{\pi}{4}\sqrt{N}\rfloor$ be the maximal number of
single runs of Grover's algorithm for which quantum searching is faster than
classical.

We compute $\pmin$ minimal value of success probability of single run of Grover's
algorithm for which we obtain a~valid result with confidence $C$
\begin{equation}
	\pmin = \min_{p}{\left\{1-(1-p)^k\geq C\right\}}.
\end{equation}

Numerically obtained values of $\pmin$ for confidence level $C=0.95$ for Grover's
algorithm are listed in Tab.~\ref{tab:grover:noise}.
\begin{table}[h!]
   \begin{center}
   \begin{tabular}{c|r|r@{.}l}
    Size of the system & \multicolumn{1}{c|}{k} & \multicolumn{2}{c}{$\pmin$}  \\
    \hline
		$N=2^3$ & 1 & 0&95000 \\
		$N=2^4$ & 2 & 0&77639 \\
		$N=2^5$ & 3 & 0&63160 \\
		$N=2^6$ & 5 & 0&45072 \\
		$N=2^7$ & 7 & 0&34816 \\
		$N=2^8$ & 10 & 0&25887 
   \end{tabular}
   \end{center}
   \caption{Values of $k$ and $\pmin$ for Grover's algorithm.}
   \label{tab:grover:noise}
\end{table}

For our numerical experiment we assume that sought element $\xi$ lies in the
``middle''  of the space of elements \textit{i.e.} $\xi=2^{n-1}$.

\begin{figure*}
\includegraphics[width=\columnwidth]{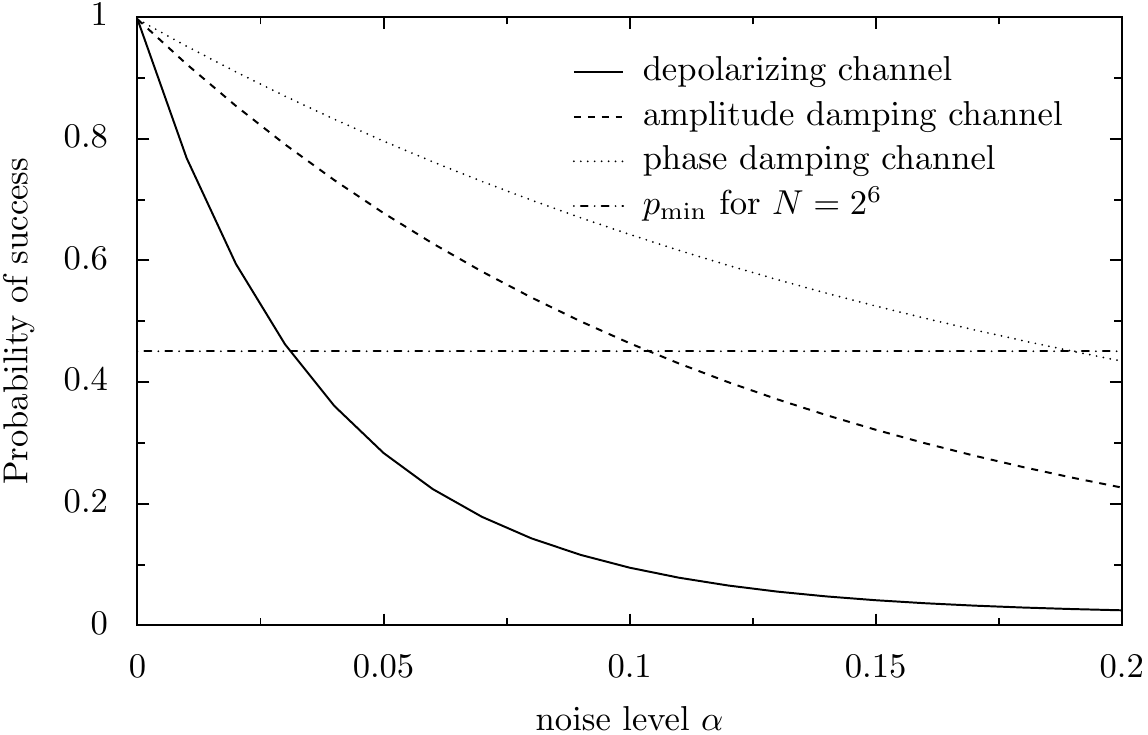}
\includegraphics[width=\columnwidth]{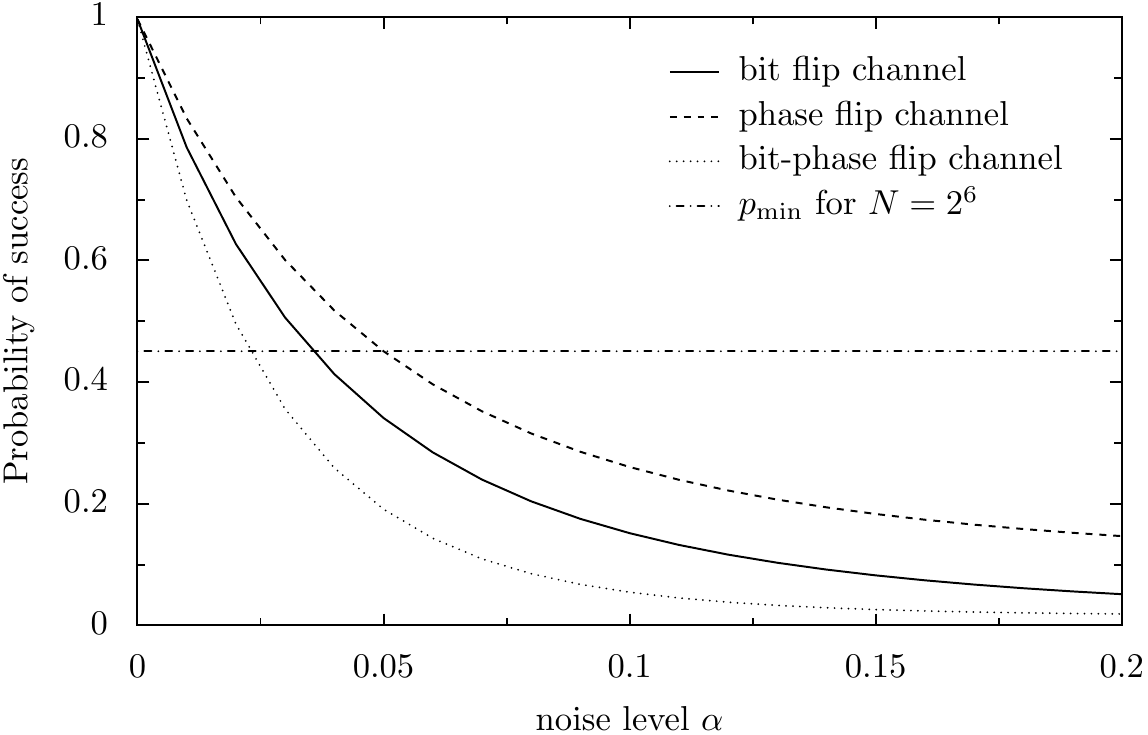}
\caption{Probabilities of successful run of Grover's algorithm in function of
the noise parameter $\alpha$. Case of six qubits. The value for which the plots
attain $\pmin$ threshold is shown in Tab.~\ref{tab:grover:noise}.}
\label{fig:grover:channels}
\end{figure*}

Plots in Fig.~\ref{fig:grover:channels} depict the influence of noise parameter
$\alpha$ on a~successful run of Grover's algorithm acting on six qubits. These
values of parameter $\alpha$ for which the plots are above threshold level
$\pmin$ can be considered as the amounts of noise which do not make quantum search
algorithm less efficient than classical search algorithms.

We can compare the probabilities from plots in Fig.~\ref{fig:grover:channels} and
these for other sizes of quantum registers with $\pmin$ and find the value of
the noise parameter $\alpha$ for which it is equal to~$\pmin$. The results of
the comparison are collected in Tab.~\ref{tab:grover:noise-params} for
confidence level $C=0.95$ and for the channels we have described in Section~\ref{sec:noisemodel}.

\begin{table}[h!]
\begin{center}
\begin{footnotesize}
\begin{tabular}{c|r@{.}l|r@{.}l|r@{.}l}
$C=0.95$ &	\multicolumn{2}{c|}{depolarising} &\multicolumn{2}{c|}{amplitude damping} &\multicolumn{2}{c}{phase damping} \\
\hline
$N=2^4$ & 0&025 & 0&069 & 0&177 \\ 
$N=2^5$ & 0&032 & 0&010 & 0&204 \\ 
$N=2^6$ & 0&031 & 0&104 & 0&190 \\ 
$N=2^7$ & 0&026 & 0&094 & 0&158 \\ 
$N=2^8$ & 0&020 & 0&075 & 0&122 \\ 
\hline
\hline
  & \multicolumn{2}{c|}{bit flip} &\multicolumn{2}{c|}{phase flip} &\multicolumn{2}{c}{bit-phase flip} \\
\hline
$N=2^4$ & 0&025 & 0&047 & 0&018 \\
$N=2^5$ & 0&032 & 0&054 & 0&024 \\
$N=2^6$ & 0&031 & 0&050 & 0&023 \\
$N=2^7$ & 0&026 & 0&041 & 0&020 \\
$N=2^8$ & 0&020 & 0&031 & 0&015 

\end{tabular}
\end{footnotesize}
\end{center}
\caption{The maximal values of noise parameter $\alpha$ for which Grover's search
algorithm is as efficient as classical search algorithm in terms of number of
uses of the oracle.}
\label{tab:grover:noise-params}
\end{table}

In the case of three qubits we have found that, if we expect confidence level $C=0.95$
or higher, Grover's algorithm is never better than classical search algorithm. It
means that if we want to get the result with high probability we need to repeat
the quantum search so many times that it is more efficient to perform this task
classically.

In other cases we have obtained the values of the noise parameter $\alpha$ between
$\sim0.010$ and~$\sim0.2$ depending on the noise type and the size of the system. 
We observe that even if the amount of noise is larger in bigger systems (what
causes the algorithm to be less efficient) the noise is compensated by the quantum
speed-up.

The results gathered in Tab.~\ref{tab:grover:noise-params} do not form a monotonic
pattern. To understand this fact we have to take into account that two factors
influence these numbers. The first one is due to the fact that the same value of
noise parameter $\alpha$ has larger influence on the quantum system for bigger
numbers of qubits and for larger $N$ the number of Grover iterations and noisy
channel applications $k$ raises. At the same time the more qubits are used to
perform the search algorithm the more important the quantum speed-up is.

\section{Summary}
\label{sec:summary}
In this work we have shown that a~new way of analysing the influence of
quantum noise on the quantum search algorithm. Our method uses the model of density
matrices and quantum channels represented in Kraus form.

We can conclude that the simulations and analysis have shown that only for small
amounts of noise the quantum search algorithm is still more efficient than any
classical algorithm.

From our numerical results we conclude that different forms of noise have
different impact on the efficiency of the quantum search algorithm. The least
destructive form of noise is phase damping, more destructive is amplitude
damping and the most destructive is the depolarizing channel.

Further work would have to take into account quantum error correcting codes and
more precise noise models dependent on the implementation. One of the research
directions would be to analyse the quantum search algorithm in the framework of
control Hamiltonians  taking into account Markovian approximation of quantum
noise.

\section{Acknowledgements}
We acknowledge the financial support by the Polish Ministry of Science and
Higher Education (MNiSW) under the grant numbers N N519 442339 and N N516 481840.
Work of P.G. was partially supported by MNiSW project number IP2010 009770.
The numerical calculations presented in this work were performed on
the \texttt{Leming} server of The Institute of Theoretical and Applied
Informatics, Polish Academy of Sciences.

\bibliography{../noisy-grover-algorithm}

\end{document}